# Interfacial Phonon Scattering and Transmission Loss in >1 μm Thick Silicon-on-Insulator Thin Films

Puqing Jiang,[1, a] Lucas Lindsay,[2] Xi Huang,[1] and Yee Kan Koh[1,b]

[1]*Department of Mechanical Engineering, National University of Singapore, Singapore 117576*

[2]*Materials Science and Technology Division, Oak Ridge National Laboratory, Oak Ridge, Tennessee 37831, USA*

## ABSTRACT

Scattering of phonons at boundaries of a crystal (grains, surfaces, or solid/solid interfaces) is characterized by the phonon wavelength, the angle of incidence, and the interface roughness, as historically evaluated using a specularity parameter $p$ formulated by Ziman [J. M. Ziman, Electrons and Phonons (Clarendon Press, Oxford, 1960)]. This parameter was initially defined to determine the probability of a phonon specularly reflecting or diffusely scattering from the rough *surface* of a material. The validity of Ziman's theory as extended to solid/solid interfaces has not been previously validated. To better understand the interfacial scattering of phonons and to test the validity of Ziman's theory, we precisely measured the in-plane thermal conductivity of a series of Si films in silicon-on-insulator (SOI) wafers by time-domain thermoreflectance (TDTR) for a Si film thickness range of 1 – 10 μm and a temperature range of 100 – 300 K. The Si/SiO$_2$ interface roughness was determined to be 0.11±0.04 nm using transmission electron microscopy (TEM). Furthermore, we compared our in-plane thermal conductivity measurements to theoretical calculations that combine first-principles phonon transport with Ziman's theory. Calculations using Ziman's specularity parameter significantly overestimate values from the TDTR measurements. We attribute this discrepancy to phonon transmission through the solid/solid interface into the substrate, which is not accounted for by Ziman's theory for surfaces. We derive a simple expression for the specularity parameter at solid/amorphous interfaces, and achieve good agreement between calculations and measurement values.

---

[a] Current address: Department of Mechanical Engineering, University of Colorado Boulder, CO 80309, USA

[b] Author to whom correspondence should be addressed. Electronic mail: mpekyk@nus.edu.sg

*This manuscript has been authored by UT-Battelle, LLC under Contract No. DE-AC05-00OR22725 with the U.S. Department of Energy. The United States Government retains and the publisher, by accepting the article for publication, acknowledges that the United States Government retains a non-exclusive, paid-up, irrevocable, world-wide license to publish or reproduce the published form of this manuscript, or allow others to do so, for United States Government purposes. The Department of Energy will provide public access to these results of federally sponsored research in accordance with the DOE Public Access Plan(http://energy.gov/downloads/doe-public-access-plan).*

# I. INTRODUCTION

Understanding the effects of boundary scattering on phonon transport is critically important for designs of nanostructured thermoelectrics[1,2] and thermal management in nanoscale electronic devices.[3,4] The lower limit to thermal conductivity via boundary scattering resistance has been historically known as the "Casimir limit", in which the phonons are 100% diffusely scattered at the boundary.[5] However, phonons can also be specularly reflected or transmitted,[6,7] especially when phonon wavelengths are long compared to the surface roughness, or when phonons are propagating at a glancing angle to the surface. The specular or diffuse nature of phonon scattering can be conveniently accounted for using a specularity parameter $p$ (a multiplicative factor to the boundary scattering rate), first introduced by Fuchs[8] in 1938: $p = 0$ corresponds to totally diffuse scattering, $p = 1$ corresponds to perfect specular reflection, and $0 < p < 1$ corresponds to partially diffuse scattering of phonons at the surface. Ziman[9] offered a simple analytical expression of $p$ as a function of the phonon wavelength $\lambda$, surface roughness $\eta$, and the incidence angle $\theta$ as follows. (Note that here we use the correct expression amended by Maznev.[10])

$$p = \exp\left(-16\frac{\pi^2}{\lambda^2}\eta^2 \cos^2\theta\right) \quad (1)$$

While Ziman's formula was initially derived for solid/air boundaries, the expression has been widely used to describe scattering of phonons at solid/solid interfaces without regarding the possibility of phonon transmission across solid/solid interfaces.[11-14] For example, Goodson and his coworkers[11,15-17] measured the in-plane thermal conductivity of high-quality Si films in silicon-on-insulator (SOI) wafers, with film thicknesses from 20 nm up to 1.6 μm. They found that the measurements were well described by classical size effects assuming fully diffuse boundaries at temperatures even as low as 20 K. They thus suspected that even high-quality SOI films have sufficient roughness to cause fully diffuse scattering. However, the roughness of Si/SiO$_2$ interfaces in SOI wafers, as revealed by TEM images from the literature[18] and our current work, is only ~0.1 nm, much shorter than the wavelengths of the dominant heat-carrying phonons in Si (~1 nm). Furthermore, in experimental studies of nanostructures, such as nanosheets and nanowires, where the boundaries were thought to be solid/air interfaces, the boundaries are actually solid/solid/air interfaces due to the presence of native oxide layers on the samples, and phonon transmission across these may have non-negligible effects on thermal transport. For example, Hertzberg et al.[19] studied the surface scattering of phonons in the frequency range 90 – 870 GHz in Si nanosheets with thickness 120 – 380 nm and roughness ~1 nm by employing a microscale phonon spectrometer technique to measure phonon transmission. Their measurements are consistent with a Monte Carlo simulation assuming 100% diffuse scattering by the surfaces, in contrast to the prediction from Ziman's specularity theory (Eq. (1)) that only <40% of the modes should be diffusely scattered. A possible reason for the discrepancy between their measurements and Ziman's theory is that the native oxide layer (solid/solid interface) is not accounted for in Ziman's specularity.

To test the validity of Ziman's expression for boundary scattering of phonons by solid/solid interfaces, here we systematically measure the in-plane thermal conductivity ($\Lambda_{\text{in}}$) of crystalline Si films with thickness $1 \leq h_f \leq 10$ μm in SOI wafers at temperatures $100 \leq T \leq 300$ K, using time-domain thermoreflectance (TDTR). We choose the film thickness to be in the range of 1 to 10 μm so that the films are both sufficiently thick allowing most of the long-wavelength phonons (which are specularly reflected according to Ziman's formula) contribute to the thermal conductivity measurement, and at the same time sufficiently thin that boundary scattering still provides a measurable resistance compared to bulk thermal conductivity of Si. By measuring the thermal conductivity at different temperatures, we can investigate phonons of varying wavelengths that dominantly contribute to the thermal conductivity. We compare our measured $\Lambda_{in}$ of Si films to first principles predictions for two boundary scattering cases: (i) Ziman's specularity equation (Eq. (1)) using the actual surface topography from transmission electron microscopy (TEM), and (ii) a simple expression we propose considering transmission of phonons at solid/amorphous interfaces. Our $\Lambda_{in}$ measurements along with the literature data agree very well with calculations using the simple expression including phonon transmission, suggesting that Ziman's theory may not be directly applicable to solid/solid interfaces as phonon transmission is not properly accounted for.

## II. MEASUREMENTS OF IN-PLANE THERMAL CONDUCTIVITY OF SI FILMS

Our samples are commercially available p-type <100> silicon-on-insulator (SOI) wafers with thickness of the device layer in the range 1−10 μm. The 1 μm thick SOI wafer was prepared by the Smart Cut$^{TM}$ process, while other SOIs were prepared by the bonding and etch-back process.[20] The device layers are single-crystalline with resistivity of 10−20 Ω cm. The dopant concentration estimated from this resistivity value is $\sim 1 \times 10^{15}$ cm$^{-3}$. We have verified that such a low concentration of impurities has a negligible effect on the lattice thermal conductivity of Si, see Ref.[21] for more discussion. To prepare the samples for TDTR measurements, we first etched away the native silicon oxide of the SOIs using hydrofluoric acid (HF), and immediately deposited a ~100 nm thick Al film as a metal transducer for the TDTR measurements. We measured the thicknesses of the Al, Si and $SiO_2$ layers by picosecond acoustics.[22]

We characterize the $Si/SiO_2$ interface roughness of our SOI wafers from TEM images; see Figure 1(c) for an example cross-section TEM image of our 5-μm-thick SOI wafer. We digitize the TEM image, and hence the interface, as a function of x- and y-coordinates. From this image, the interface roughness of $\eta = 0.11 \pm 0.04$ nm and the correlation length of $\xi = 2.2 \pm 0.7$ nm are determined by analyzing the extracted function of the interface by a MATLAB code, see Supplementary Information for more details. The roughness of the $Si/SiO_2$ interfaces in our SOI wafers is similar to that obtained from TEM images of SOI wafers previously.[18]

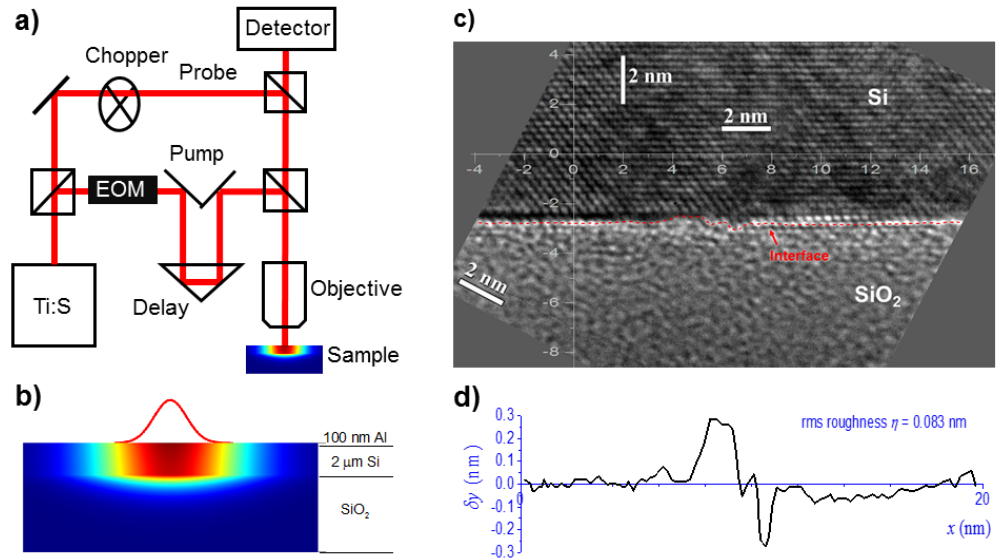

Figure 1. (a) Schematic of TDTR setup to measure the in-plane thermal conductivity of Si films; (b) Temperature profile in SOI during TDTR measurements. (c) TEM image of the cross-section of our 5-μm-thick SOI wafer; (d) Digitized $Si/SiO_2$ interface for roughness estimation. We obtained the RMS roughness $\eta$ as $0.11 \pm 0.04$ nm for the $Si/SiO_2$ interface of our SOI wafers.

We employ time-domain thermoreflectance (TDTR)[23-25] to measure the in-plane thermal conductivities of the Si films, see Figure 1 (a) for a schematic of our TDTR setup.[24] In TDTR measurements, a train of sub-picosecond laser pulses is split into a pump beam and a probe beam. The pump beam, modulated by an electro-optic modulator, is absorbed by the transducer of the sample and periodically heats the sample at a modulation frequency $f$. The periodic temperature response at $f$ at the sample surface is then monitored by a synchronized but time-delayed probe beam via thermoreflectance, using a photodetector and a lock-in amplifier. We then extract the thermal conductivity of the Si films by comparing the ratio of the in-phase and out-of-phase signals of the lock-in amplifier at $f$, $R_f = -V_{in} / V_{out}$, to calculations from a diffusive thermal model.[23]

We note that the thermal conductivity of the Si films is mainly derived from the *out-of-phase* and not the in-phase signals of TDTR measurements.[26] The in-phase TDTR signals mainly correspond to temperature decay due to heating of a single laser pulse. The relaxation time of the in-phase signals is mostly determined by the thermal conductance of the Al/Si interface. On the other hand, the out-of-phase TDTR signals are essentially the out-of-phase temperature response due to periodic heating at the modulation frequency $f$. With the periodic heating, heat diffuses a distance $d_p$ into the surface; $d_p = \sqrt{\Lambda/\pi C f}$ is called the thermal penetration depth, with $C$ the volumetric heat capacity.[27] Thus, TDTR measurements are only sensitive to the thermal properties of the sample within a distance $d_p$.

To measure $\Lambda_{in}$ of Si thin films using TDTR, we used a small $1/e^2$ radius of $w_0$ = 5.5 μm for the laser beams and a low modulation frequency of $f$ = 0.5 MHz to achieve an uncertainty ~10%. At such a low modulation frequency, the thermal penetration depth $d_p$ in Si is ~7 μm at 300 K and ~23 μm at 100 K. With $d_p > w_0$, heat transfer is generally three-dimensional.[28] However, since the thermal resistance of the SiO$_2$ substrate is high, heat flows primarily in the in-plane direction in our SOIs, as illustrated in Figure 1(b). Therefore, the measured temperature response at the surface is sensitive mostly to the in-plane thermal conductivity and not the cross-plane thermal conductivity, see the sensitivity plots of our measurements in the Supplementary Figure S2. When the Si films are relatively thick (e.g., >10 μm), the measured surface temperature response is slightly sensitive to the cross-plane thermal conductivity the Si films. For those cases, we use the cross-plane thermal conductivity we previously reported from separate TDTR measurements[29] in our thermal model. The uncertainty of the cross-plane thermal conductivity is ~10%.

We extract both the effective thermal conductivity of the Si films and the Al/Si interface conductance $G$ simultaneously because they affect the TDTR signals in different manners in our measurements. Our measured Al/Si interface conductance $G$ (=320 MW m$^{-2}$ K$^{-1}$ at 300 K) is independent of the Si film thickness and is consistent with literature values.[30] Our measurements are not sensitive to the thermal conductivity of the Al transducer and the Si/SiO$_2$ interface conductance, see the sensitivity plots in the Supplementary Figure S2. We varied the Si/SiO$_2$ interface conductance from 30 MW m$^{-2}$ K$^{-1}$ to 30 GW m$^{-2}$ K$^{-1}$ in the thermal model, which varied the effective thermal conductivity of the Si films by < 2%. We also carefully choose appropriate laser spot sizes and modulation frequencies such that our TDTR measurements are not affected by the thermal conductivity of the underlying SiO$_2$ layer, see the sensitivity plots in the Supplementary Figure S2. Due to the high in-plane thermal conductivity of the Si films, heat more readily dissipates sideway through the Si films than across the high resistant SiO$_2$.

We verified that our measurements of $\Lambda_{in}$ were not affected by the laser spot size or the modulation frequency used in our TDTR measurements. Previous measurements of bulk Si using TDTR showed a spot size dependence.[30,31] This dependence should not be as pronounced in Si thin films as the mean-free-paths of low-frequency phonons are limited by the film thickness $h_f$ due to boundary scattering at the interfaces, thus ballistic effects should be minimal. We tested this assertion by measuring the $\Lambda_{in}$ of the Si films at different temperatures using different spot sizes $w_0 =$ 5.5 μm, 11 μm, and 27 μm, with $f$ fixed at 0.5 MHz. We observe no significant spot size dependence for the measured $\Lambda_{in}$ when $f$ is sufficiently low (e.g., 0.5 MHz), see Figure 2. To check the frequency dependence, we measured the thermal conductivity of bulk Si as a function of modulation frequency, see our data previously reported in Ref [29]. No frequency dependence was observed in bulk Si at room temperature and only slight frequency dependence was observed at 100 K, which is consistent with previous TDTR measurements of bulk Si in the literature.[27,32] We note that the thermal penetration depth $d_p$ for bulk Si at $f =$ 0.5 MHz and 100 K is 12 μm, much larger than the film thickness of our samples. Thus, we deduce that our measured $\Lambda_{in}$ of Si films are not affected by frequency dependent artifacts.

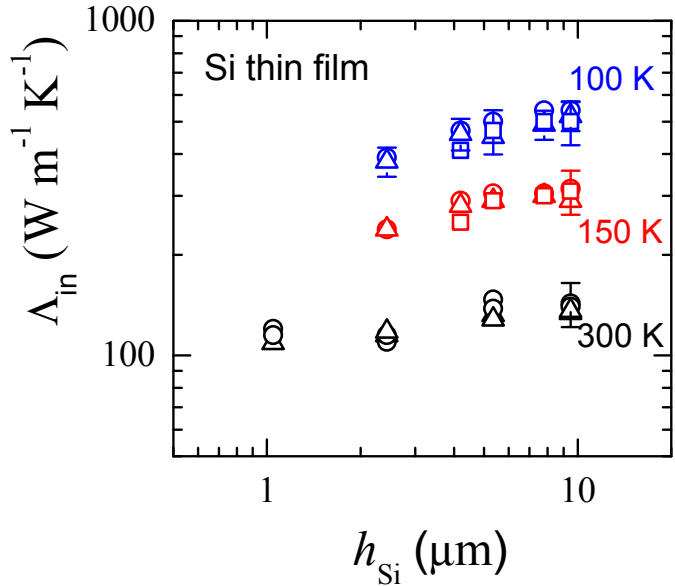

Figure 2. In-plane thermal conductivity of Si films as a function of film thickness at 300 K, 150 K, and 100 K, measured by TDTR using different laser spot sizes (circles: $w_0 = 5.5$ μm, triangles: $w_0 = 11$ μm, and squares: $w_0 = 27$ μm) at 0.5 MHz.

## III. CALCULATIONS OF SPECULAR SCATTERING OF PHONONS IN SI FILMS

We employed a Peierls-Boltzmann transport methodology for theoretical modeling of phonon conduction within the relaxation time approximation, which is sufficient in determining thermal conductivity of silicon because of the strong non-conserving phonon scattering in silicon.[33] The harmonic interatomic force constants (IFCs), which are necessary to derive the phonon dispersions and phonon relaxation times, were calculated from density functional perturbation theory. More details of the IFC calculations are given in Ref.[34]. The lattice thermal conductivity of bulk Si can be calculated using the formula:[35]

$$\Lambda = \frac{1}{3V} \sum_{(\vec{q},j)} \frac{\hbar^2 \omega(\vec{q},j)^2}{k_B T^2} n_0(\vec{q},j)\left(n_0(\vec{q},j)+1\right) v(\vec{q},j)^2 \tau_c(\vec{q},j) \qquad (2)$$

In this expression, $V$ is the crystal volume, $(\vec{q}, j)$ designates a phonon with wavevector $\vec{q}$ in branch $j$, $\hbar$ is the Planck constant, $\omega(\vec{q}, j)$ is the phonon frequency, $k_B$ is the Boltzmann constant, $n_0(\vec{q}, j)$ is the Bose distribution function, $v(\vec{q}, j)$ is the phonon velocity, and $\tau_c(\vec{q}, j)$ is the combined phonon relaxation time. There are several mechanisms for phonon scattering, and the Matthiessen's rule is used to sum up these effects as $\tau_c(\vec{q}, j)^{-1} = \tau_N(\vec{q}, j)^{-1} + \tau_U(\vec{q}, j)^{-1} + \tau_I(\vec{q}, j)^{-1}$, with $\tau_N(\vec{q}, j)$, $\tau_U(\vec{q}, j)$ and $\tau_I(\vec{q}, j)$ the relaxation times for the normal phonon-phonon scattering processes, the umklapp phonon-phonon scattering processes, and the phonon-isotope scattering, respectively. These relaxation times are determined from quantum perturbation theory and have been described in detail previously.[34,36] Note that the frequencies, velocities, and relaxation times are all wavevector and branch dependent.

From the relaxation times of the bulk phonons $\tau_{c,\text{bulk}}$ we derive the relaxation times of phonons in thin films $\tau_{c,\text{film}}$ for in-plane heat conduction using the Fuchs-Sondheimer relationship[8,37] as:

$$\frac{\tau_{c,\text{film}}}{\tau_{c,\text{bulk}}} = 1 - \frac{3}{2}\text{Kn}(1-p)\int_1^\infty \left(\frac{1}{t^3} - \frac{1}{t^5}\right)\frac{1-e^{-t/\text{Kn}}}{1-pe^{-t/\text{Kn}}}dt \qquad (3)$$

The Knudsen number is defined as $\text{Kn} = v\tau_{c,\text{bulk}}/h_f$, with $h_f$ the thickness of the film, and $p$ is the specularity parameter. Note that the parameter $p$ is explicitly calculated by Eq. (1) for each mode and thus is a function of both phonon wavelength $\lambda = 2\pi/q$ and the cosine of the incidence angle, $\cos\theta = q_z/q$, where $q_z$ is the wavevector component perpendicular to the surfaces and $q$ is the wavevector magnitude.

The relaxation time of Si thin films $\tau_{c,\text{film}}$ is then plug into Eq. (2) to calculate the in-plane thermal conductivity of Si thin films. By varying the specularity parameter *p* and comparing the model calculations with our experimental measurements, we can determine the role of specular scattering in SOI wafers and test whether the Ziman theory is still valid in our experimental conditions.

## IV. RESULTS AND DISCUSSION

We present our measurements of the in-plane thermal conductivity of Si films $\Lambda_{in}$ along with some other literature measurements of Si thin films[17,38] in Figure 3 as a function of film thickness $h_f$ at different temperatures of 300 K, 150 K, and 100 K, and compared with our model calculations. Examination of Figure 3 indicates that predictions of $\Lambda_{in}$ of Si films using the specularity parameter based on Ziman's equation and the actual interface roughness deviate significantly from the measurements, whereas predictions assuming purely diffusive interfaces ($p = 0$) agree well with the experiments over a broad range of film thicknesses from 20 nm to 20 µm. The good agreement between our measurements and the predictions assuming fully diffuse scattering is rather surprising considering the very smooth interfaces revealed by the TEM images, but the same conclusion was reached in previous studies of Si thin films.[11,15]

A possible explanation for the discrepancy between our measured data and Ziman's prediction is that it was originally derived for solid/air interfaces (surfaces) for which all phonons are either specularly reflected ($p = 1$) or diffusely scattered ($p = 0$). At solid/solid interfaces, however, phonons can also be transmitted. To account for phonon transmission across solid/solid interfaces, we derive a simple expression for the specularity parameter $p = (1 - \alpha)p_{Ziman}$ where $p_{Ziman}$ is from Ziman's theory (Eq. 1) and $\alpha$ is a mode dependent probability of phonon transmission from the Si to the SiO$_2$ substrate. We propose that the underlying amorphous SiO$_2$ layer destroys the phase coherence of the transmitted phonons before the phonons are re-emitted back into the crystalline Si layer. Hence, the overall effect is similar to diffuse scattering of phonons at the interfaces, and the transmitted phonons can thus be considered as diffusely scattered at the interfaces, though with a physically different picture.

To test the proposed expression for *p*, we attempt two transmission probability $\alpha$ profiles for the Si/SiO$_2$ interfaces. First, we estimate a frequency-independent $\alpha$ from the diffuse mismatch model (DMM); $\alpha = I_2/(I_1 + I_2)$, with $I_i = \sum_j v_{i,j}^{-2}$, where $i = 1$ and 2 stand for Si and SiO$_2$, respectively, and $v_j$ is the sound velocity of phonons of polarization *j*, see Table 1 for the summary of sound velocities in crystalline Si and amorphous SiO$_2$ we used. Second, we use the frequency-dependent transmission probability profile across the Si/SiO$_2$/Al interface that was recently derived from TDTR measurements.[39] (The SiO$_2$ at the interface represents the native oxide on Si surfaces.) The experimentally derived $\alpha$ is the lower limit of phonon transmission at the Si/SiO$_2$ interface, as we assume that Al/SiO$_2$ and SiO$_2$/Si interfaces are decoupled. We plot the DMM and the experimental $\alpha$ in Figure 4, as a function of phonon frequency.

Table 1. Sound velocities $v_L$ and $v_T$ of single-crystalline Si and amorphous SiO$_2$ determined from their literature measurements of elastic constants.[40,41]

|  | $\rho$ (kg m$^{-3}$) | $v_L$ (m s$^{-1}$) | $v_T$ (m s$^{-1}$) |
|---|---|---|---|
| Si | 2331 | 8474 | 5843 |
| SiO$_2$ | 2200 | 5953 | 3743 |

In Figure 3, we compare our measurements to calculations using the proposed expression, with $\alpha$ either estimated from DMM or derived from prior experiments. We find that despite the different transmission probabilities of the two cases, both predictions of $\Lambda_{in}$ of Si films compare well with our measurements and the predictions assuming fully diffusive interfaces. To understand these results, we further compare the phonon transmission probability $\alpha$ with the specularity parameter as a function of phonon frequency in Figure 4. We find that Ziman's theory predicts that low-frequency phonons of <3 THz (which carry significant amount of the heat, ~50% and ~65% in bulk Si at 300 K and 100 K, respectively) should be specularly reflected at the interfaces, resulting in little loss in the phonon energy. These low-frequency phonons, however, will transmit into the amorphous $SiO_2$ layer instead due to the similar acoustic impedances of Si and $SiO_2$. As a result of the transmission and subsequent scattering of phonons in amorphous $SiO_2$, a significant amount of phonon energy is lost and the thermal conductivities of Si films is substantially reduced.

Our research has a broad impact on nanoscale heat transfer research as the presence of native oxide layers on Si is ubiquitous in experimental conditions. This work explains why past experiments on Si nanostructures, either suspended thin films,[38] nanosheets,[19] or nanowires,[42] all agree well with the model predictions assuming fully diffuse boundary scattering. Our conclusion is also consistent with the results of a molecular dynamics simulation,[43] which predicts that the presence of a 1-nm-thick amorphous layer (either Si or silica) would reduce the thermal conductivity of a 15-nm-diameter single crystalline Si nanowire by more than 70%, from 45 W m$^{-1}$ K$^{-1}$ to 12.5 W m$^{-1}$ K$^{-1}$.

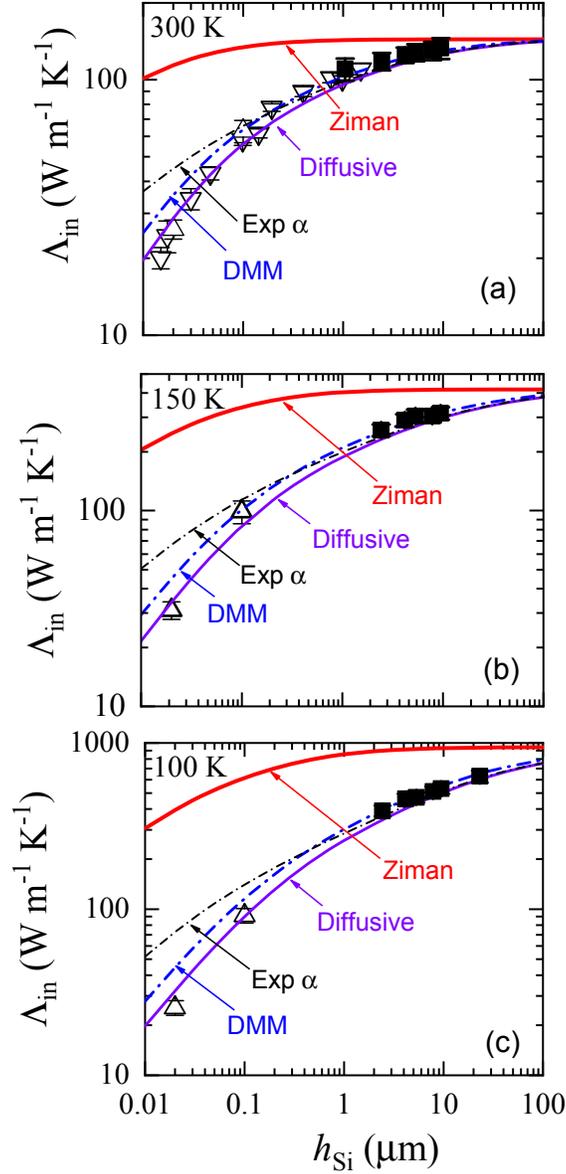

Figure 3. In-plane thermal conductivity of silicon thin films at (a) 300 K, (b) 150 K, and (c) 100 K. The solid symbols are the current measurements and the open symbols are similar measurements from the literature.[17,38] Curves are from our first-principles calculations coupled with different specularity models: the one labeled "Diffusive" assumes totally diffuse scattering ($p = 0$) at the interface, while the one labeled "Ziman" uses Ziman's equation (Eq. 1) to calculate the specularity parameter $p_{Ziman}$, with interface roughness $\eta = 0.11$ nm determined from TEM images of the interface, and the other two labeled "DMM" and "Exp $\alpha$" take into account phonon

transmission in the specularity parameter as $p = (1 - \alpha)p_{\text{Ziman}}$, with $\alpha$ being the transmission coefficient determined using DMM and from experimental measurements of phonon transmission across Si/SiO$_2$/Al, respectively.

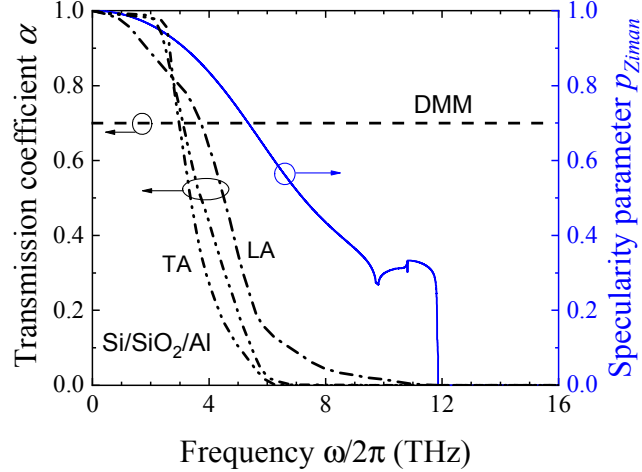

Figure 4. Transmission coefficient of phonons at the Si/SiO$_2$ interface estimated using DMM (dashed lines), along with the literature measurements of phonon transmission across the Si/SiO$_2$/Al interfaces (dash-dot line),[39] which serves as a lower limit of phonon transmission from Si to SiO$_2$, for comparison. The specularity parameter calculated using Ziman's formula (solid line) is also included.

## V. CONCLUSION

In summary, we studied the validity of Ziman's theory for specular scattering of phonons at interfaces by carefully measuring the in-plane thermal conductivity of Si films in SOI wafers over a thickness range of 1 – 10 μm at temperatures 100 – 300 K. Our measured in-plane thermal conductivity values of Si films, while in agreement with other experimental data in the literature, deviate significantly from the model predictions using the specularity parameter based on Ziman's theory. The discrepancy between Ziman's theory and our measurements can be explained by the transmission of phonons across the Si/SiO$_2$ interface. This is not considered in Ziman's theory which was derived for phonon scattering at surfaces (solid/air interfaces), though it was extensively used in the literature to describe solid/solid interfaces. Since the underlying amorphous SiO$_2$ layer destroys the phase coherence of the transmitted phonons, the effect of the transmission of phonons at Si/SiO$_2$ interfaces is equivalent to diffuse scattering of those phonons, though the physical picture is different. We thus propose a simple expression for the specularity parameter that takes into consideration transmission of phonons across solid/amorphous interfaces. Calculations using this expression agree well with our measured data. This work sheds light on the role of amorphous layers on the scattering of phonons at solid/amorphous interfaces, prevalent in a host of microelectronics applications and fundamental nanoscale research.

## ACKNOWLEDGEMENTS

This work was supported by the Singapore Ministry of Education Academic Research Fund Tier 2 under Award No. MOE2013-T2-2-147 and Singapore Ministry of Education Academic Research Fund Tier 1 FRC Project FY2016. L. L. acknowledges support from the Laboratory Directed Research and Development Program of Oak Ridge National Laboratory, managed by UT-Battelle, LLC, for the U. S. Department of Energy.